# On the Half-Life of $^{71}$Ge and the Gallium Anomaly

E. B. Norman[1], A. Drobizhev[2], N. Gharibyan[3], K. E. Gregorich[3], Yu. G. Kolomensky[1,2], B. N. Sammis[3], N. D. Scielzo[3], J. A. Shusterman[3], and K. J. Thomas[3]

[1.] University of California, Berkeley, CA 94720 USA
[2.] Lawrence Berkeley National Laboratory, Berkeley, CA 94720 USA
[3.] Lawrence Livermore National Laboratory, Livermore, CA 94550 USA

## Abstract

Recent discussions about the origin of the so-called Gallium Anomaly have motivated a remeasurement of the half-life of $^{71}$Ge. We have conducted three separate measurements using dedicated planar Ge detectors – one with $^{55}$Fe as a standard, one with $^{57}$Co as a standard, and one stand-alone $^{71}$Ge measurement. Our results yield a half-life of 11.468 ± 0.008 days, which is consistent with, but significantly more precise than, the currently-accepted value. With this experiment, the potential explanation of the Gallium Anomaly being due to an unexpectedly long $^{71}$Ge half-life has been ruled out, leaving the anomaly's origin as an open question.

The Gallium Anomaly is based on the fact that several experiments have reported lower than expected ($\nu_e$,$e^-$) reaction rates on $^{71}$Ga [1–4]. The vast majority of these reactions are thought to involve transitions between the ground states of $^{71}$Ga and $^{71}$Ge. The origin of this discrepancy is not understood, having first been observed in the SAGE [2,5,6] and GALLEX [7,8] experiments, and confirmed with stronger statistical significance recently by the BEST experiment [4]. Given that the predicted reaction rates are believed to be on firm footing, this deficit potentially is an indication of new neutrino physics [2,9]. The cross section for such reactions can be deduced from the electron-capture decay half-life, $t_{1/2}$, of $^{71}$Ge. The most recent evaluated value gave $t_{1/2}$=11.43±0.03 days [10], based on measurements done in 1985 [11]. Recently, it was suggested that a somewhat larger value of 12.5 – 13.5 days could reduce or eliminate this anomaly [12] and that careful new measurements of this half-life were warranted [13]. To thoroughly address this issue, we have conducted three independent, high-precision measurements of the $^{71}$Ge half-life.

We produced $^{71}$Ge via neutron activation. Two separate Ge targets were irradiated at the TRIGA reactor of the McClellan Nuclear Research Center (MNRC) to produce three distinct counting samples. The first irradiation target was 10.6 mg of natural-abundance germanium oxide $GeO_2$, obtained from Fairmont Chemical Co. sealed in a quartz ampule and irradiated for 8 hours at 250 kW in the central core of the reactor. The irradiated sample was transported to the Lawrence Livermore National Laboratory (LLNL) and opened in water to prevent contamination from the dispersible $GeO_2$ powder. The water sealed the openings of the two ampule half-pieces, allowing for the transfer of irradiated $GeO_2$ powder with a few drops of ethanol using a fine-tip glass pipette. The ethanol-$GeO_2$ slurry was placed on a plastic disk and dried slowly under a heat lamp. The dry powder was then taped down against the plastic disk with 1-mil thick Kapton tape. An additional 0.0254 mm thick Kapton tape was affixed over the entire disk to ensure sample integrity for an extended period of time. Two such counting samples were prepared from the irradiated $GeO_2$ powder. The second irradiation target consisted of 0.82 g of detector-grade high-purity Ge (1.4 – 1.5 ×$10^{10}$ cm$^{-3}$ net p-type) procured from Mirion Technologies Inc. This was cut into three small pieces of approximately 3 mm × 3 mm × 5 mm dimensions and the surface cleaned via aggressive etching in a 3:1 $HNO_3$:HF mixture, subsequently quenched in deionized water and rinsed with methanol. This material was irradiated in the graphite reflector of the MNRC reactor for 19 minutes at 800 kW. Two of these Ge pieces were mounted together on a plastic disk and held in place with two 1-mil thick pieces of Kapton tape. Each of the three irradiated samples were counted on coaxial detectors at the LLNL Nuclear Counting Facility to assess radiopurity. Initially, the $^{71}$Ge activities were accompanied by a few short-lived isotopes, such as $^{77}$Ge ($t_{1/2}$=11.3 hr). The three $^{71}$Ge sources were stored for 14 days to allow these short-lived activities to decay away before the half-life measurements began.

The decay of $^{71}$Ge to $^{71}$Ga by electron capture was monitored by the pair of K-shell x-ray lines at 9.225 keV/9.252 keV (with total intensity 39.4%) and 10.26 keV/10.264 keV (with total intensity 4.9%) on three planar high-purity germanium (HPGe) detectors at the LLNL Nuclear Counting Facility. Two were ORTEC detectors with crystal diameters of 36 mm and having thin Be windows. The third was a Canberra detector with crystal diameter of 52 mm and having a thin carbon composite window. The energy resolutions of the three detectors were all approximately

0.35 keV (FWHM) at 5.9 keV. Each detector was housed in its own commercial low-background graded shield.

We performed one absolute $^{71}$Ge half-life measurement and two relative measurements using radioactive standards. We mounted one of the GeO$_2$ samples approximately 15.4 cm away from the front face of one of the ORTEC detectors. The other GeO$_2$ sample was mounted together with a weak source of $^{57}$Co ($t_{1/2}$ = 271.74 ± 0.06 days) near the front face of the other ORTEC detector. The Ge wafer pieces were placed together with a weak $^{55}$Fe source ($t_{1/2}$ = 2.744 ± 0.009 years) near the front face of the Canberra detector [14,15]. Data were collected from near zero to approximately 300 keV and binned in 4096 channels in 1-day-long time intervals using three separate ORTEC DSPEC data acquisition systems. Energy spectra were stored at the end of each day for offline analysis. Over an 80-day counting period, the shields were never opened and the sources were never moved. Figure 1 illustrates the relevant portions of the spectra we obtained from each of the three detector systems.

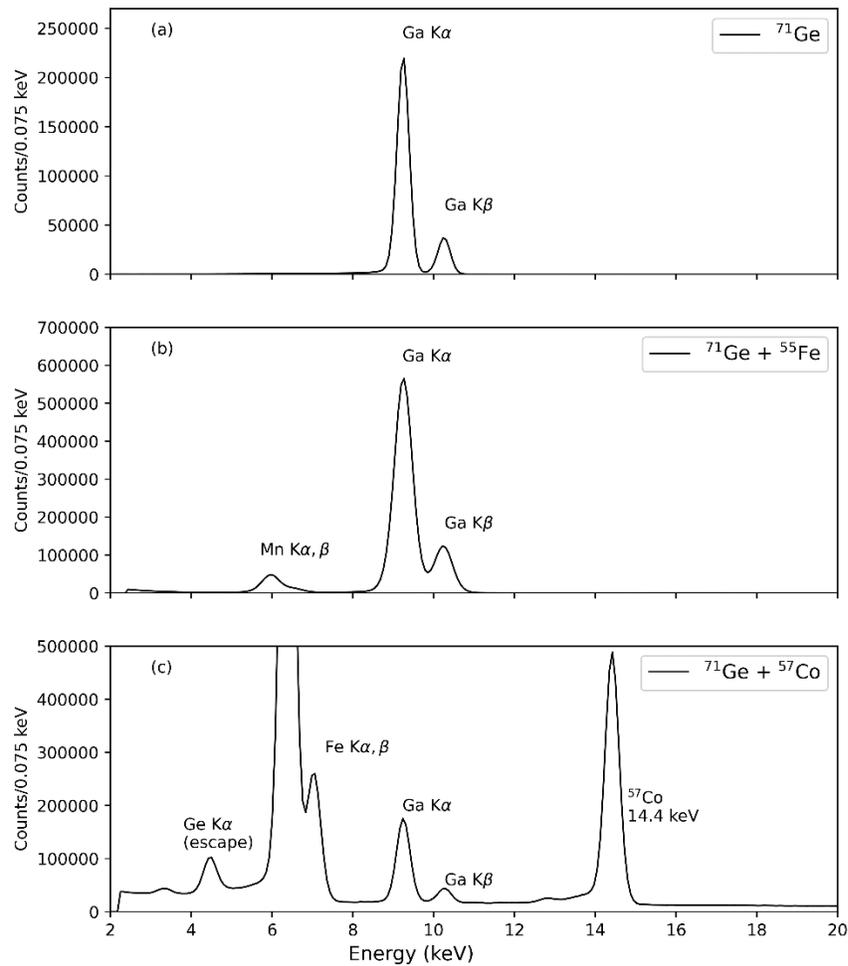

*Figure 1: Relevant portions of the spectra obtained on day 20 from the three measurements we performed: (top) $^{71}$Ge alone, (middle) $^{71}$Ge+$^{55}$Fe, (bottom) $^{71}$Ge and $^{57}$Co. Each spectrum represents 24 hours of counting.*

From the 80 spectra collected from each of the three detector systems, we utilized two different methods to extract peak areas of the Ga x rays and in the cases of the relative measurements, those of the Mn K-shell x rays ($^{55}$Fe) and the 14-keV γ ray ($^{57}$Co). In the first method (Analysis 1), the counts in peak and background windows in each spectrum were integrated and the net peak areas and their uncertainties were then determined by subtraction. To minimize the large uncertainties Analysis 1 suffered as a result of signal interference between $^{71}$Ge, $^{55}$Fe, and $^{57}$Co activities, this analysis uses only data collected between Day 49 and Day 60 ($^{71}$Ge+$^{55}$Fe exposure) and between Day 1 and Day 30 ($^{71}$Ge+$^{57}$Co exposure).

In the second method (Analysis 2), portions of the spectra containing the peaks of interest were fit using a peak-shape parameterization similar to that suggested by Campbell and Maxwell [16]. The main part of the peak is a Gaussian detector-resolution function,

$$G(x) = H_G \exp\left(\frac{-(x-\mu)^2}{2\sigma^2}\right),$$

where $x$ is the channel number, $H_G$ is the Gaussian amplitude, $\mu$ is the peak centroid, and σ is the standard deviation. An exponential tail on the low energy edge of the peak (convoluted with the Gaussian detector resolution) is introduced to account for charge trapping in detector defects. The peak described by these functions sits on a smoothly varying background. There are several background function options: none, constant, linear, quadratic, or exponential, selected based on the location of the peak-free background region. A step in the background occurs at $\mu$, due to several effects, including Compton electrons escaping the active detector volume, and coherent scattering in place of Compton scattering. This background step is also convoluted with the Gaussian detector resolution.

Dead-time corrections, ranging from 0.33% at the start of data taking to 0.14% at the end, were applied to the data obtained from the stand-alone $^{71}$Ge sample. For the $^{55}$Fe + $^{71}$Ge data, we analyzed the ratio of the Ga and Mn x rays. For the $^{57}$Co + $^{71}$Ge data, we analyzed the ratio of the Ga x rays to the 14-keV γ ray. By using such ratios, the effects of possible variations in detector and/or electronics performance are largely cancelled out. In the course of examining the $^{57}$Co + $^{71}$Ge data, we discovered relatively weak Ge K-shell x rays within our analysis windows that appear to have been produced by 14-, 122-, and 136-keV γ rays interacting with the dead layers of our HPGe detectors. After our half-life measurements were completed and the $^{71}$Ge sample was removed, we used the same detector to obtain spectra from just the $^{57}$Co source. This data was then used in both of our analysis methods to correct for the contributions of the Ge x rays to our data. Decay curves from our three measurements and both analyses are shown in Figure 2.

The decay curves observed from each detector were fit using a linear least squares technique. The data from the stand-alone $^{71}$Ge source provides a direct measure of the $^{71}$Ge half-life and are well described by an exponential function. In the cases of the relative measurements, a fit to the ratio of activities between $^{71}$Ge and a standard ($^{55}$Fe or $^{57}$Co) provides a measure of the difference in decay rates between $^{71}$Ge and the that of the standard. The data are also well described by the exponential function, except for the $^{57}$Co + $^{71}$Ge sample in Analysis 2, where a constant term is needed to account for the uncertainty in the $^{57}$Co background subtraction. From the fitted difference in decay rates, we compute the $^{71}$Ge half-life using the accepted values of the half-lives of the standards [14,15]. The results of our three measurements are shown in Table 1. In cases where the goodness of fit metric $\chi^2$/d.f. deviates significantly from unity indicating non-statistical fluctuations, the uncertainties have been determined by multiplying the fitted statistical uncertainties by the square root of $\chi^2$/d.f. The relatively large $\chi^2$/d.f. in Analysis 2 fits for the $^{71}$Ge + $^{55}$Fe and $^{71}$Ge + $^{57}$Co arise from 1) the scatter in the decay curves for the Mn x ray and the 14.4-keV $\gamma$ ray, and 2) the interference from the $^{55}$Fe and $^{57}$Co in the Ga x-ray peak-fitting regions. The effect from the latter is visible in the residual plots of Figure 2.

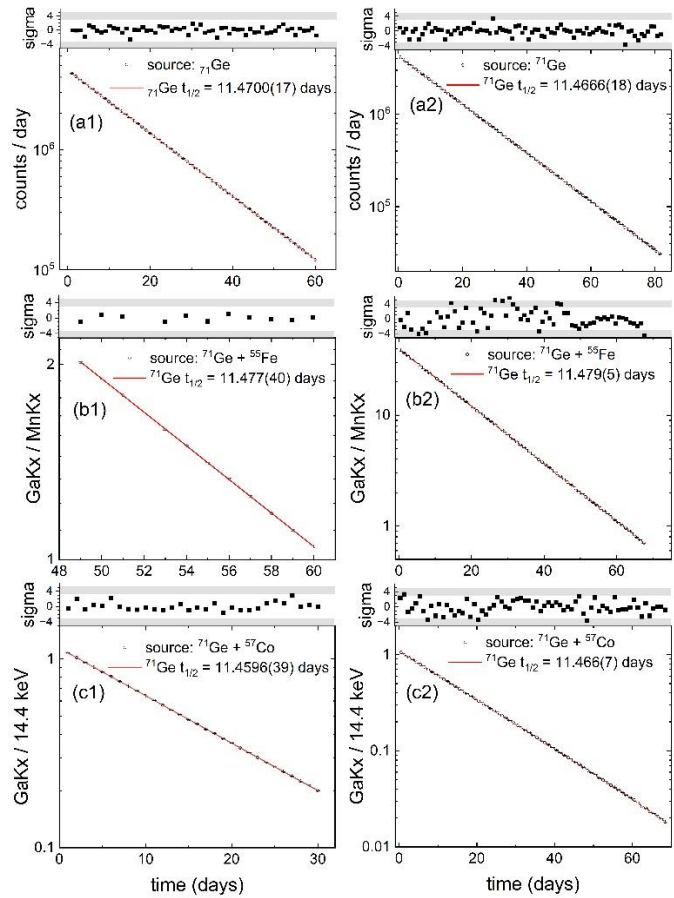

*Figure 2: Decay curves observed from our three independent measurements of the $^{71}$Ge half-life. (top) Ga K-shell x rays as a function of time from the stand-alone $^{71}$Ge source. (middle) the ratio of the Ga K-shell x rays to the Mn K-shell x rays from the $^{71}$Ge + $^{55}$Fe sample (c) ratio of the Ga K-shell x rays to the 14-keV $\gamma$ ray from the $^{71}$Ge + $^{57}$Co sample. Figures on left are from Analysis 1 and figures on right are from Analysis 2.*

Table 1: Results of the present measurements of the half-life of $^{71}$Ge.

| Sample | Analysis 1 | | Analysis 2 | |
|---|---|---|---|---|
| | Half-life (days) | $\chi^2$/d.f. | Half-life (days) | $\chi^2$/d.f. |
| $^{71}$Ge | 11.4700±0.0017 | 49.1/58 | 11.4666±0.0018 | 110.7/80 |
| $^{71}$Ge + $^{55}$Fe | 11.477±0.040 | 5.3/9 | 11.479±0.005 | 339.6/61 |
| $^{71}$Ge + $^{57}$Co | 11.4596±0.0039 | 30.8/28 | 11.466±0.007 | 180.5/66 |

The two most precise measurements are from the detector exposed only to the $^{71}$Ge source. While the two analysis techniques determine the peak areas differently, the resulting half-life values are highly correlated. We determine the value of the $^{71}$Ge half-life as an arithmetic average of two results without adjusting the uncertainty:

$$t_{1/2}(^{71}\text{Ge}) = 11.4683 \pm 0.0017 \text{ (stat.) days.}$$

Systematics in the measurement can come from the precision of the detector clock, time-dependent detection efficiency (e.g. due to rate-dependent deadtime corrections), and pathologies in background subtraction. The measurements performed with the standard sources allow us to monitor and calibrate out time-dependent effects; moreover, most time-dependent effects cancel in the ratio of the $^{71}$Ge and a standard decay rate. The background-subtraction systematics are amplified in the ratio analysis due to the non-trivial signal interference between the $^{71}$Ge, $^{55}$Fe, and $^{57}$Co activities. Therefore, we use the results from the $^{71}$Ge+$^{55}$Fe and $^{71}$Ge+$^{57}$Co samples to estimate the systematic uncertainties.

First, we fit the decay rate curves for the $^{55}$Fe ($^{55}$Mn x-ray peaks in Fig. 1b) and $^{57}$Co (14.4-keV line in Fig. 1c) samples to a combination of two exponentials: the primary function with the free decay rate, and the subdominant function with the half-life fixed to our measured value of 11.4683 days to account for spurious contamination of the main peak by the $^{71}$Ge sample. The fits are shown in Fig. 3, and result in the estimates of the $t_{1/2}(^{55}\text{Fe})= 2.98 \pm 0.21$ (stat.) yr and $t_{1/2}(^{57}\text{Co})= 270.7 \pm 1.9$ (stat.) days. These are in excellent agreement with canonical values [14,15]. If we were to treat the statistical uncertainty on $t_{1/2}(^{57}\text{Co})$ as a gauge of the size of any potential rate-dependent systematic skewing of our measured half-life from the actual value, it would only contribute ±0.0034 days to $t_{1/2}(^{71}\text{Ge})$, owing to the significantly shorter half-life of $^{71}$Ge.

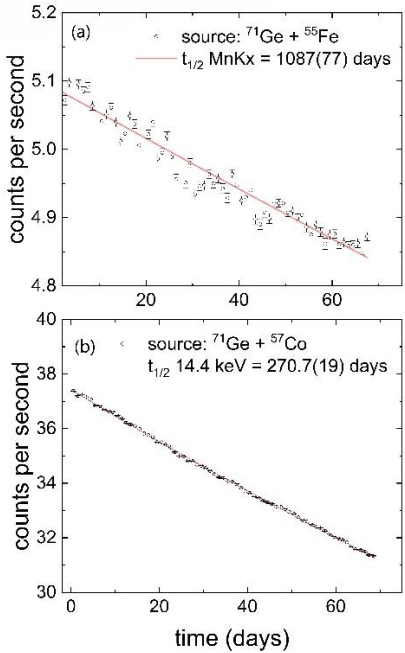

Figure 3: Time dependence of the decay rates for the (a) $^{55}$Fe and (b) $^{57}$Co datasets.

Instead, we considered the spread of the values of $t_{1/2}(^{71}\text{Ge})$ present between our six results presented in Table I as a more conservative estimate the systematic uncertainty. The standard deviation of these six values is 0.0082 days; it captures the background-subtraction systematics of the $^{71}$Ge+$^{55}$Fe and $^{71}$Ge+$^{57}$Co samples, and also covers, by construction, rate-dependent effects.

Finally, combining the statistical and systematic uncertainties, we determine the half-life of $^{71}$Ge to be

$$t_{1/2}(^{71}\text{Ge}) = 11.4683 \pm 0.0017\ (\text{stat.}) \pm 0.0082\ (\text{syst.}) = 11.468 \pm 0.008\ \text{days}.$$

Figure 4 compares the present results to those from previous measurements of the $^{71}$Ge half-life.

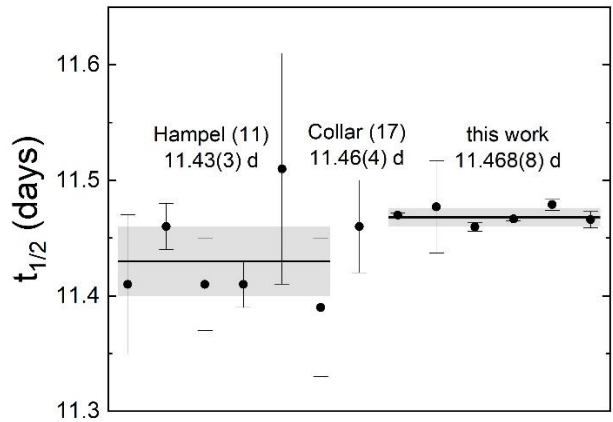

Figure 4: Results from all reported measurements of the $^{71}$Ge half-life since 1985.

Our final result for the half-life of $^{71}$Ge is in very good agreement with the accepted value of 11.43 ± 0.03 days [10] and agrees with, but is significantly more precise, than the recently reported 11.46 ± 0.04 day half-life [17]. Thus, as pointed out in a new detailed paper [18], the origin of the Gallium Anomaly remains an open question.

## Acknowledgements

We thank Wesley Frey and the staff of the MNRC for carrying out the neutron irradiations that produced the $^{71}$Ge used in our measurements. This material is based upon work supported by the US Department of Energy (DOE) Office of Science under Contract No. DE-AC02-05CH11231, and by the DOE Office of Science, Office of Nuclear Physics under Contract No. DE-FG02-08ER41551. This work was performed in part under the auspices of the U.S. Department of Energy by LLNL under Contract DE-AC52-07NA27344. This work was partially supported by the Laboratory Directed Research and Development Program at LLNL under project tracking code 23-SI-004.

## References

[1] J. N. Abdurashitov et al., *Measurement of the Response of a Gallium Metal Solar Neutrino Experiment to Neutrinos from a $^{51}$Cr Source*, Physical Review C **59**, (1999).
[2] J. N. Abdurashitov et al., *Measurement of the Response of a Ga Solar Neutrino Experiment to Neutrinos from a $^{37}$Ar Source*, Physical Review C **73**, (2006).
[3] F. Kaether, W. Hampel, G. Heusser, J. Kiko, and T. Kirsten, *Reanalysis of the Gallex Solar Neutrino Flux and Source Experiments*, Physics Letters B **685**, 47 (2010).
[4] V. V. Barinov et al., *Results from the Baksan Experiment on Sterile Transitions (BEST)*, Physical Review Letters **128**, (2022).
[5] C. Giunti and M. Laveder, *Short-Baseline Active-Sterile Neutrino Oscillations?*, Mod. Phys. Lett. A **22**, 2499 (2007).
[6] C. Giunti and M. Laveder, *Statistical Significance of the Gallium Anomaly*, Physical Review C **83**, (2011).
[7] P. Anselmann and GALLEX, *First Results from the $^{51}$Cr Neutrino Source Experiment with the GALLEX Detector*, Physics Letters B **342**, 440 (1995).
[8] W. Hampel et al., *Final Results of the $^{51}$Cr Neutrino Source Experiments in GALLEX*, Physical Review B **420**, 114 (1998).
[9] M. Laveder, *Unbound Neutrino Roadmaps*, Nuclear Physics B **168**, 344 (2007).
[10] B. Singh and J. Chen, *Nuclear Structure and Decay Data for A=71 Isobars*, Nuclear Data Sheets **188**, (2023).
[11] W. Hampel and L. P. Remsberg, *Half-Life of $^{71}$Ge*, Physical Review C **31**, 666 (1985).
[12] C. Giunti, Y. F. Li, C. A. Ternes, and Z. Xin, *Inspection of the Detection Cross Section Dependence of the Gallium Anomaly*, Physics Letters B **842**, (2023).
[13] S. R. Elliott, V. Gavrin, and W. Haxton, *The Gallium Anomaly*, ArXiv:2306.03299v1 (2023).
[14] M. R. Bhat, *Nuclear Data Sheets for A=57\**, Nuclear Data Sheets **85**, (1998).
[15] H. Junde, *Nuclear Data Sheets for A = 55*, Nuclear Data Sheets **109**, 787 (2008).

[16] J. L. Campbell and J. A. Maxwell, *A Cautionary Note on the Use of the Hypermet Tailing Function in X-Ray Spectrometry with Si(Li) Detectors*, Nuclear Instruments and Methods in Physics Research B **129**, 297 (1997).
[17] J. I. Collar and S. G. Yoon, *New Measurements of $^{71}$Ge Decay: Impact on the Gallium Anomaly*, Physical Review C **108**, (2023).
[18] S. R. Elliott, V. N. Gavrin, W. C. Haxton, T. V. Ibragimova, and E. J. Rule, *Gallium Neutrino Absorption Cross Section and Its Uncertainty*, Physical Review C **108**, (2023).